\documentclass[10pt,letterpaper]{article}
\usepackage[top=0.85in,left=0.75in,footskip=0.75in,marginparwidth=2in]{geometry}

\usepackage[utf8]{inputenc}

\usepackage{cite}
\usepackage{upgreek}
\usepackage[caption=false,font=footnotesize]{subfig}
\usepackage{graphicx}
\usepackage{upgreek}
\usepackage{booktabs} 
\usepackage[table]{xcolor} 

\usepackage{nameref,hyperref}

\usepackage[right]{lineno}

\usepackage{microtype}
\DisableLigatures[f]{encoding = *, family = * }

\raggedright
\usepackage{changepage}

\usepackage[aboveskip=1pt,labelfont=bf,labelsep=period,singlelinecheck=off]{caption}

\makeatletter
\renewcommand{\@biblabel}[1]{\quad#1.}
\makeatother

\usepackage{lastpage,fancyhdr,graphicx}
\usepackage{epstopdf}
\fancyheadoffset[L]{2.25in}
\fancyfootoffset[L]{2.25in}

\usepackage{color}

\definecolor{Gray}{gray}{.25}

\usepackage{graphicx}

\usepackage{sidecap}

\usepackage{wrapfig}
\usepackage[pscoord]{eso-pic}
\usepackage[fulladjust]{marginnote}
\reversemarginpar

\begin{document}
\vspace*{0.35in}

\begin{flushleft}
{\Large
\textbf\newline{First Attempts in Automated Defect Recognition in Superconducting Radio-Frequency Cavities}
}
\newline
\\
Author Marc Wenskat{1},

\bigskip
\bf{1} Deutsches Elektronen-Synchrotron, 22607 Hamburg, Germany
\\
\bf{2} Universit\"at Hamburg, 22607 Hamburg, Germamy
\\
\bigskip
* marc.wenskat@desy.de

\end{flushleft}

\section*{Abstract}
The inner surface of superconducting cavities plays a crucial role to achieve highest accelerating fields. The industrial fabrication of cavities for the European X-Ray Free Electron Laser (EXFEL) and the International Linear Collider (ILC) HiGrade Research Project allowed for an investigation of this interplay with a large sample on different cavities undergoing a standardized procedure. For the serial inspection of the inner surface, the optical inspection robot OBACHT was constructed and to analyze the large amount of data, represented in the images of the inner surface, an image processing and analysis code was developed. New variables to describe the cavity surface were obtained. Two approaches using these variables and images to automatically detect defects has been implemented and tested. In addition, a decision-tree based approach of classifying defect free surfaces regarding their accelerating performance was tested and found to be physically valid.

\section{Introduction}

Superconducting niobium radio-frequency cavities, with the TESLA shape, are fundamental for accelerators like the European X-Ray Free Electron Laser (EXFEL), the International Linear Collider (ILC), the European Spallation Source (ESS) or the Linac Coherent Light Source II (LCLS-II) \cite{XFEL_TDR, ILC_TDR, ESSTDR, LCLS-II}. 
To utilize the operational advantages of superconducting cavities, the inner surface has to fulfill quite demanding requirements. 
 Electromagnetic RF fields, used for the particle acceleration, penetrate the inner surface of the superconducting resonator. The penetration depth is about 40 nm (London penetration depth for niobium) and the near-surface composition and surface topography play a crucial role for the final performance. For an investigation of the behavior of cavities under such RF fields, an optical surface inspection tool was developed at KEK and Kyoto University, the "Kyoto camera system" \cite{Iwashita2008,Tajima2008}. The intention is, that an optical inspection of the inner surface, which is exposed to the RF field, leads to a better understanding of limitations observed during RF tests. While a general correlation was found between low field quenches (transition from superconducting to normal-conducting phase) and localized defects seen in optical inspections \cite{Watanabe,Watanabe2008,Moller2009,Geng2009a,Aderhold2010b,Singer2010}, an automated detection of these defects and a possible classification has not been achieved yet. First attempts in an automated classification of defects but also defect free surfaces are presented in this work. In addition, a correlation of surface structures of defect free cavities and their RF performance was done.

\section{Optical Inspection and Image Processing}
During the construction of the European XFEL \cite{Reschke2017}, more than 100 TESLA cavities underwent subsequent surface treatments, acceptance tests at 2\,K, and optical inspections within the ILC-HiGrade research program \cite{CORDIS, Navitski2013,Singer2016,MW_SUST,Wenskat2015a}. The optical inspection of the inner surface of SRF cavities is a well-established tool at many laboratories \cite{Watanabe}. 
Its purpose is to characterize and understand performance limitations which lead to a breakdown of the accelerating field and to allow optical quality assurance during cavity production. Theoretical calculations have shown that accelerating fields of $50\, \mathrm{MV/m}$  are achievable if surface structures and localized defects are below 10\,$\upmu \mathrm{m}$ \cite{Sara1995,Xie2009}. With a designed accelerating field of $23.6\, \mathrm{MV/m}$ for the European XFEL and an aimed average accelerating field on the order of $35\, \mathrm{MV/m}$, the resolution of the optical inspection system should be on the level of 10\,$\upmu \mathrm{m}$ to resolve the inner surface.  

This resolution and the rather large surface of a cavity results in a large amount of images which needs to be inspected and classified. Hence, an algorithm was developed which enables an automated surface characterization. This algorithm delivers a set of optical surface properties, which describe the inner cavity surface and allow for a framework for quality assurance of the fabrication procedures. Furthermore, this framework shows promising results for a better understanding of the observed limitations in defect free cavities.
\subsection{OBACHT}
\begin{figure}[!htb]
	\centering
		\includegraphics*[width=\linewidth]{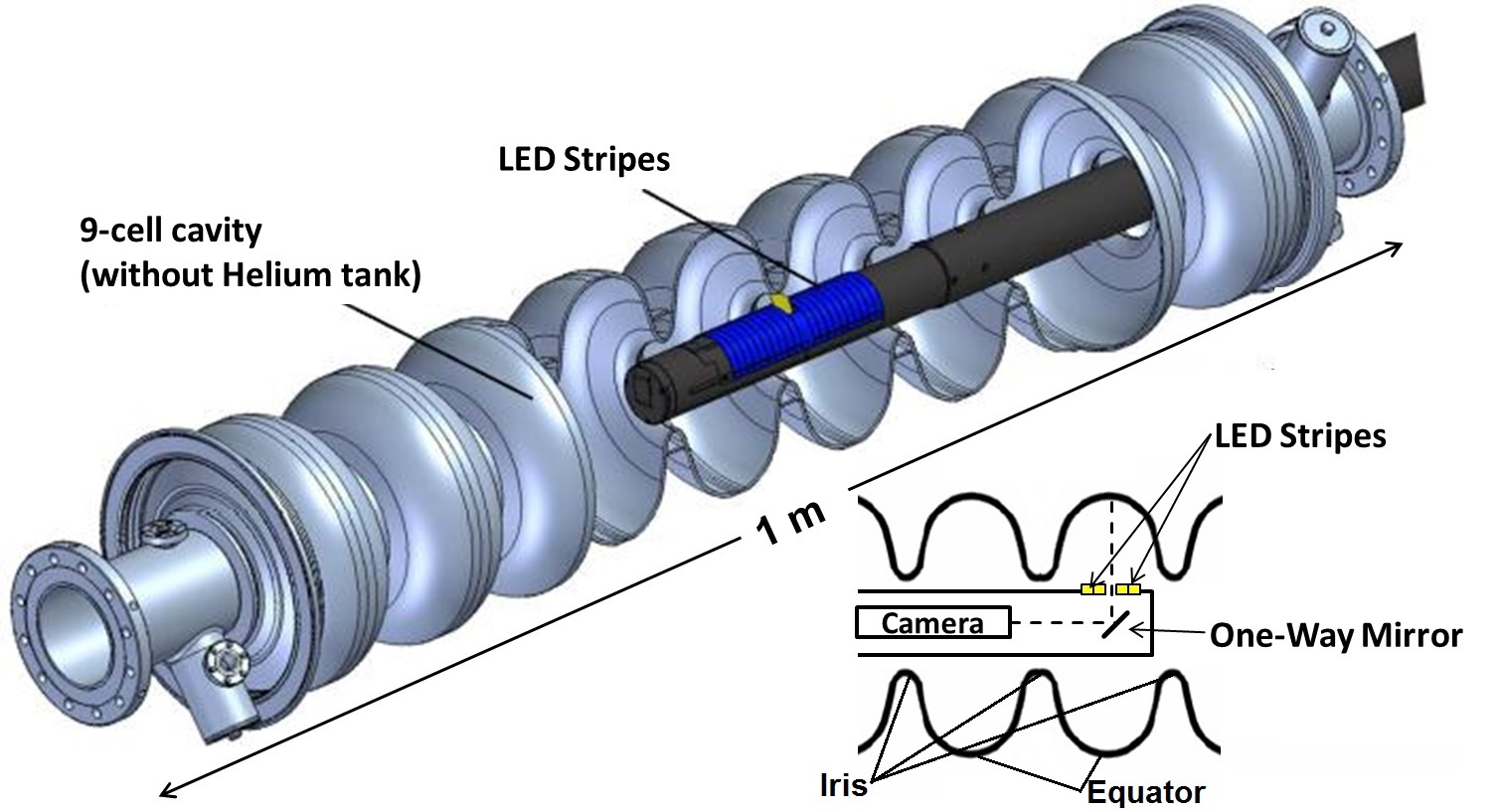}
	\caption{Schematic of the Kyoto Camera System used at DESY. The camera is viewing the inner surface via a $ 45^\mathrm{o}$-tilted one way mirror. Behind this one way mirror, three LEDs are mounted for the central illumination. $2 \times 10$ acrylic stripes with LEDs are mounted left and right from the opening in the tube for a more detailed illumination. On the lower right, a cut of the set up and the regions called equator and iris are shown.}
	\label{fig:sketch}
\end{figure}

\begin{figure}[!htb]
	\centering
		\includegraphics*[width=\linewidth]{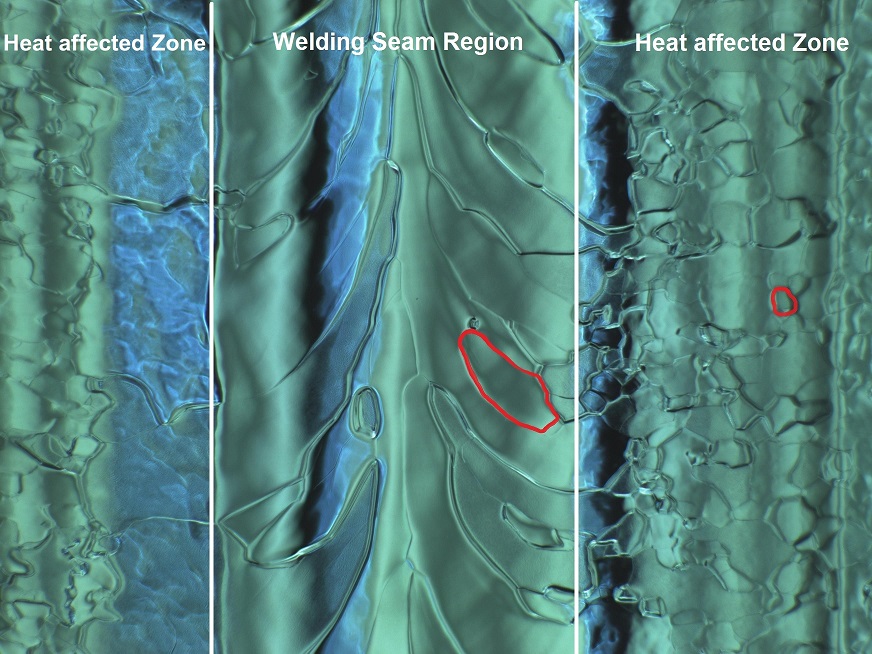}
	\caption{Image of the inner cavity surface with the equatorial welding seam in the image center taken with OBACHT. The image size is $9 \times 12\,\mathrm{mm^2}$. The red contours are examples of grain boundaries identified with the image processing algorithm in the welding seam region (WS) and the heat affected zone (HAZ).}
	\label{OBACHT}
\end{figure}

A fully automated robot for optical inspection,  the "\textbf{O}ptical \textbf{B}ench for \textbf{A}utomated \textbf{C}avity inspection with \textbf{H}igh resolution on short \textbf{T}imescales" (OBACHT), has been developed at DESY and is continuously in use since 2009. It is equipped with a high-resolution camera (Kyoto Camera System), which resolves structures down to $12\,\upmu \mathrm{m}$ for properly illuminated surfaces \cite{Iwashita2008,Tajima2008,Iwashita2009}. The details of OBACHT and the optical system are described in \cite{Lemke,Sebastian,Wenskat2015}. The optical system is fitted inside a tube which has a diameter of 50\,mm to fit into the cavity without colliding with the so called irides, where the cavity has an inner diameter of 60\,mm and are the most narrow parts of a cavity (see Fig. \ref{fig:sketch}). 
In this tube, the camera is installed altogether with a low-distortion lens. The camera system images the surface via a $ 45^\mathrm{o}$-tilted one way mirror which can be continuously adjusted to other angles in order to inspect other cavity regions apart from equator or iris. To match the focal distance of the camera to the camera -cavity surface distance, the camera is moved along the rotational axis of the rod, controlled by a motor driven lead-screw. For illumination, acrylic stripes (two Light Emitting Diodes - LED - per strip) attached to the camera tube around the camera opening are installed, together with three additional LEDs behind the one way mirror inside the camera tube.
The equatorial images are of main interest for this analysis. This is because the highest magnetic field in a cavity, and hence the possible highest ohmic losses destroying the superconducting state, are at the equatorial welding seam region including the so-called "heat affected zone" (HAZ), a region next to the remelted weld where the crystals grew due to the deposited heat during welding. Figure \ref{OBACHT} shows an image of the inner cavity surface taken with OBACHT.
With given cavity geometry and optical set up, an individual image covers $5^\mathrm{o}$ of an equator. To have a small overlap at the edges of an image, an image is taken each $4.8^\mathrm{o}$. This results in 75 images per equator and 675 equator images per cavity where the image covers an area of $9 \times 12\,\mathrm{mm^2}$ and consists of $3488 \times 2616\,\mathrm{pixel}$ per color layer. The objects of interest within an image of the inner cavity surface are grain boundaries and defects. In order to identify and quantify those objects, an image processing and analysis algorithm has been developed.

\subsection{Image Processing and Analysis}
\begin{figure}[!htb]
   \centering
   \includegraphics*[width=\linewidth]{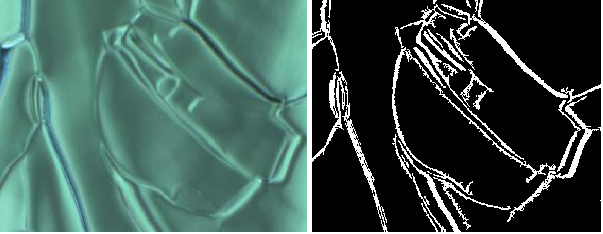}
   \caption{Left: a detail of an OBACHT image is shown. Right: the same detail after the image processing algorithm. Grain boundaries (white) are visible.}
   \label{GB_area}
\end{figure}

The main goal of the image processing algorithm is to identify grain boundaries and defects, regardless of their position within the image which shows a non-uniform illumination, as can be seen in Figure \ref{OBACHT}. The approach of this algorithm is, to apply a sequence of high-pass filter and local contrast enhancements, to project pixels which belong to grain boundaries onto a gray scale interval, which is distinct from the background. After this projection, a histogram based segmentation of the processed image is performed. The Otsu-segmentation assumes, that the image contains two classes of pixels (grain boundary and background), where the intensity values follow a bi-modal distribution, and calculates the optimum threshold separating the two classes \cite{Otsu}. 
The output is a binary image with the same size as the input image and contains grain boundary pixels - white/logical one - and background pixels -black/logical zero. As a last step of the image processing, groups of connected white pixels which form a grain boundary are classified as a single object with a connected component labeling algorithm called run-length encoding \cite{Labelling} and a labeled binary image called $L_1$ is obtained. 
An example of such a binary image is given in Figure \ref{GB_area}.
\newline
The image processing algorithm was benchmarked and tested regarding resolution accuracy. To investigate the resolution limit of the image processing algorithm, the USAF 1951 test chart was used. The smallest objects, which are still detected as individual stripes, are part of group five, element one. This results in a algorithm resolution of 15.63\,$\mathrm{\upmu m}$ with good contrast. The algorithm resolution is slightly below the resolution of the optical system (12.4\,$\mathrm{\upmu m}$). This is because of the filtering and smoothing procedure, which tends to connect objects with a distance smaller than four pixels, and due to the area cut to remove shot noise objects. Additionally, a contrast dependent resolution check was performed - for details see \cite{Wenskat2017}. As a result, it was observed that lines, spaced with a distance of 10\,pixels (35\,$\mathrm{\upmu m}$) or more - or roughly twice the algorithm resolution - have been resolved independently of the contrast. Lines with a spacing below that distance had to have an intensity difference to their background of at least 16\,Bit to be resolved. This rather robust identification regardless of the contrast of objects is based on the local contrast enhancement which is part of the algorithm.     
\newline
The image processing algorithm can be interpreted as a classifier, since the binary image classifies each pixel either as a background (no boundary) or a foreground (boundary) pixel. In order to have the ability to decide whether a pixel is rightly or wrongly classified, a test image with known properties is used. Here, the J\"ahne test image $g_1$ \cite{Jaehne2004} was used and the accuracy and the positive predictive value (PPV) of the algorithm can be calculated to be 85\,\% and 84\,\% respectively.
\newline
The aim of the image analysis is to identify features in the binary image. Those are grain boundaries with varying width which are not symmetric. Hence, it is nontrivial to define important properties like diameter, center of mass, eccentricity $\epsilon$ or orientation $\phi$ of an object. The method to overcome this problem, is to find an ellipse which has the same second central moment as the pixel distribution of the grain boundary \cite{Hu1962}. Within this ellipses approach, the grain boundary area is the total of pixels a boundary consists of. This number is retrieved from the binary image and then multiplied by the pixel size, which is a property of the very optical system. At OBACHT, this value is $12.25\,\upmu \mathrm{m}^2$. With the given resolution at OBACHT, the experimentally obtained upper relative error for the grain boundary area obtained that way is 3\,\%, similar to \cite{Patil2011}. Since the error converges with $\frac{1}{n^2}$, an area of 122 pixels at OBACHT only has a relative uncertainty of only 1\,\%. The uncertainty of the center of mass of the object is nearly constant. With the given resolution at OBACHT the uncertainty is less than ten pixels or $35\,\upmu \mathrm{m}$.

\begin{figure}[htbp]
	\centering
		\includegraphics[width=0.4\textwidth]{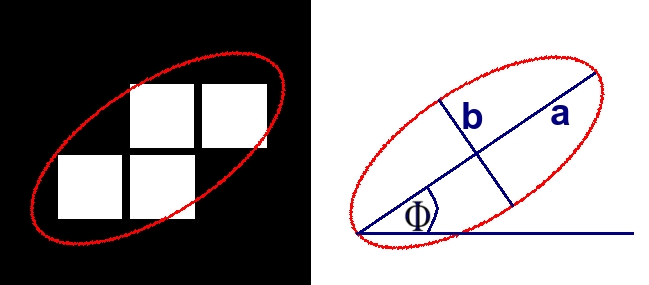}
	\caption{On the left side, a four pixel object and the ellipse with the same second central moment is shown. On the right side, the ellipse with its major axis a and minor axis b and the horizontal z-axis for the angle assignment $\Phi$ is shown.}
	\label{fig:ellipse}
\end{figure}

The orientation of an object is defined as the angle of the major axis of the ellipses with respect to an axis perpendicular to the welding seam., see figure \ref{fig:ellipse} An upper limit of the uncertainty of $5^\mathrm{o}$ is derived with \cite{Klette1999,Liao1993}. 

In order to define a figure of merit for the roughness of an object with OBACHT, two assumptions are made. The first assumption is that the intensity of the reflected light is dependent on the roughness and structures of the cavity surface. This means that a change in the intensity is either caused by a geometric gradient or a change in reflectivity. A geometric gradient exists either at a grain boundary or a defect, while a change in reflectivity can be caused by an impurity. The second assumption is that the curvature of the elliptical cavity is negligible within the studied area and surface seen by the image can be considered to be flat within the depth of field of the camera. 
In order to describe the surface roughness variations in the images obtained by the camera, we define the quantity $\mathrm{R_{dq}}$ to be the average gradient of the image intensity of an object. It is based upon measurements of the surface texture of cavity ISO 25178 \cite{iso25178}, for which a steeper increase in the surface roughness near to an edge or weld contributes to a larger average roughness value $\mathrm{R_{dq}}$.

A statistical noise arises from the Signal-to-Noise-Ratio (SNR) of the image sensor in the camera, which yields a $\delta \mathrm{R_{dq}}$ of $\frac{0.011}{\sqrt{N}} \frac{\mathrm{Bit}}{\mathrm{\upmu m}}$. A systematic uncertainty due to image focus was found to be $\frac{\delta \mathrm{R_{dq}}}{\mathrm{R_{dq}}}= 3\,\%$. For more details on the image processing algorithm and explicit definitions and discussion of the obtained variables see \cite{Wenskat2015,Wenskat2017} and for first results of surface classification and quality assurance and control during the mass fabrication see \cite{Wenskat2015a,MW_SUST}.

\section{Defect Recognition}
The biggest challenge of a defect recognition algorithm for superconducting cavities is the highly irregular surface of the grain structure and the large variety of possible defects. The fractions of known classes of defects make up roughly 40\% of the total number of observed defects but a much smaller fraction of performance relevant defects. Performance relevant defects are typical unique shaped geometric or chemical irregularities with a small sample size per class. Hence, training classifiers like the efficient Viola-Jones algorithm can not cover all relevant defects \cite{VJ1,VJ2}. Untreated cavities present a simpler surface for several kinds of image classifier, but most defects are only visible after chemical surface treatments which again pronounces the surface grain structure. This surface is difficult to illuminate homogeneously and create grains with varying sizes and topologies. To tackle these issues, two approaches were tested to identify defects. The first algorithm has no fixed assumptions about a standard surface or surface feature but compares each object with its surrounding neighbors with varying metrics based on the Mahalanobis distance \cite{Mahalanobis}. The other algorithm tries to reconstruct the image which is decomposed into small areas as a summation of eigenvectors \cite{eigenface1} and marks deviations above a defined threshold between the reproduced surface and the actual surface as a defect. 

\subsection{Object Oriented Approach}
\begin{figure}[tb]
   \centering
   \includegraphics*[width=\linewidth]{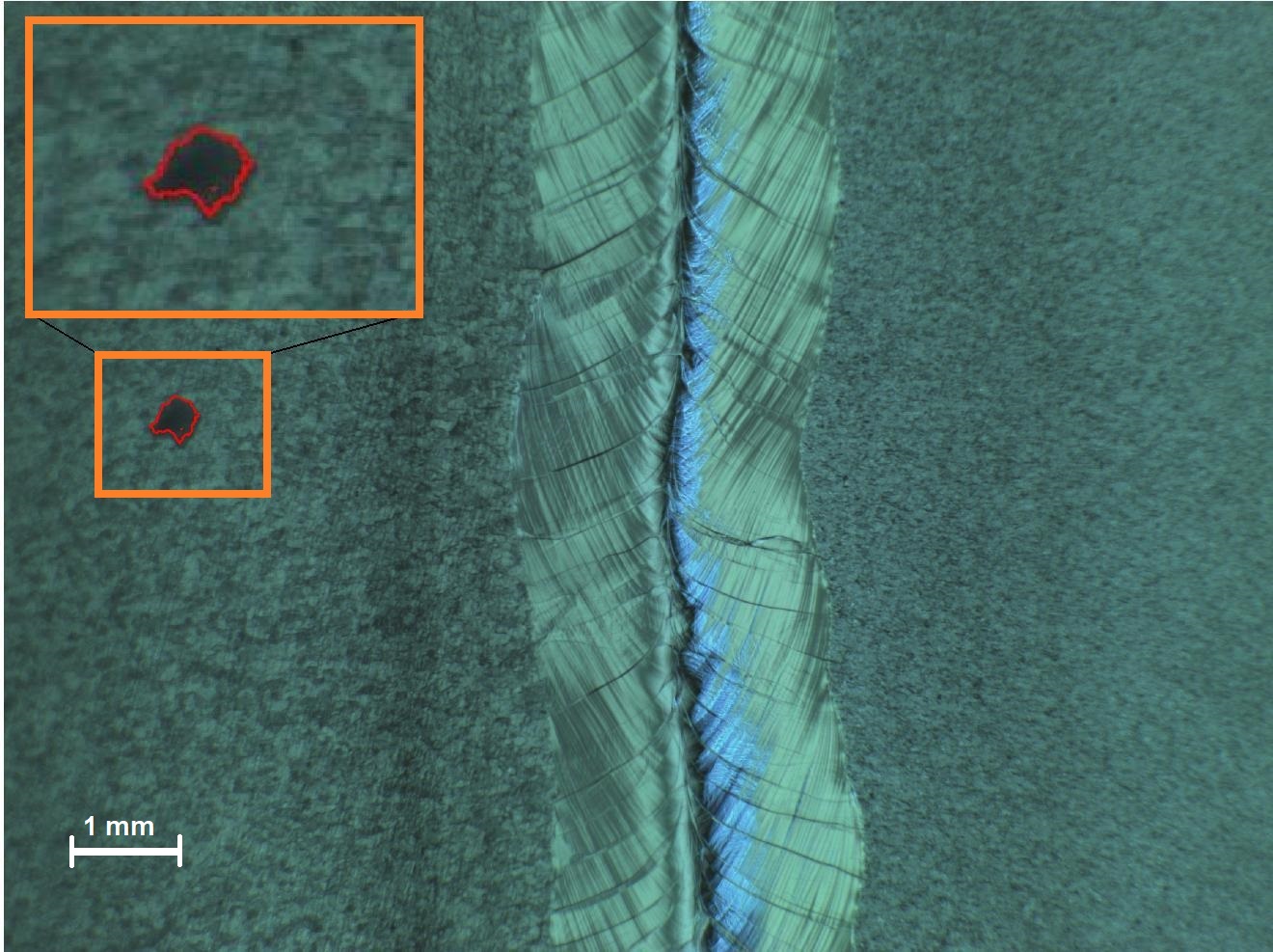}
   \caption{The dark spot - encircled by the boundary detected by the algorithm - is an aluminum contamination on an untreated surface introduced in the sheet after quality control and during rolling. This defect - after surface chemistry - reduced the quench field.}
   \label{Defectfound}
\end{figure}

On average, each image taken of the equator region and processed to obtain the labeled binary image $L_1$ contains 1700 to 2200 grain boundaries and other objects such as defects, stains or other remnants from the surface treatment or simple pollutions like fibres \cite{Wenskat2015}. Using the now defined variables, a 6-dimensional phase space is constructed where each object is represented by the six variables: area A, the ratio between the area and the perimeter f, eccentricity $\epsilon$, orientation $\phi$, the position of the object defined in the image coordinates $\vec{R} = \left(m,n\right)$ and the roughness $\mathrm{R_{dq}}$. The set of objects $\mu$ which are used to classify an object under investigation is defined as followed 
\begin{equation}
\mu = \{ \vec{x} \in L_1 | \left\| \vec{R}_{object} - \vec{R}_{\vec{x}} \right\| \leq 2 \cdot a_{object} \}
\end{equation}
with a as major axis length of the object under investigation, $\vec{R}$ the position of the object in $L_1$ and $\left\| \bullet \right\|$ the euclidean distance. This radius definition allows to adjust the set size to be studied to the object size. A fixed value would either unnecessarily increase the computational resources and mix different areas in the image (too large radius) or the set of objects is not representative (too small radius). 
Since the dimensions in the phase space have different units and the range of the values differ by several magnitudes, the euclidean distance cannot be used to quantify a distance in the phase space. Instead, the so-called Mahalanobis-Distance will be used, which is defined for two objects described by the vectors $\vec{x},\vec{y} \in \mu$  
\begin{equation}
d_M(\vec{x},\vec{y})=\sqrt{\left( \vec{y} - \vec{y} \right)^T \Sigma^{-1} \left( \vec{y} - \vec{y} \right)}
\end{equation}
where $\Sigma$ is the covariance matrix. In case that the covariance matrix equals the unit matrix, the equation simplifies to the euclidean distance. Otherwise, the correlations between different properties are taken into account, which is necessary for this case. 
An intuitive explanation for this definition is that the distance is a measure of how many standard deviations away is an object from the center of mass of the set of other objects used for comparison. The further away it is, the more likely that the object is not part of the set. 
The threshold distance used to classify an object as defect has been set to $5 \sigma$. Figure \ref{Defectfound} shows a defect found in an image. 
 
\begin{figure}[tb]
   \centering
   \includegraphics*[width=\linewidth]{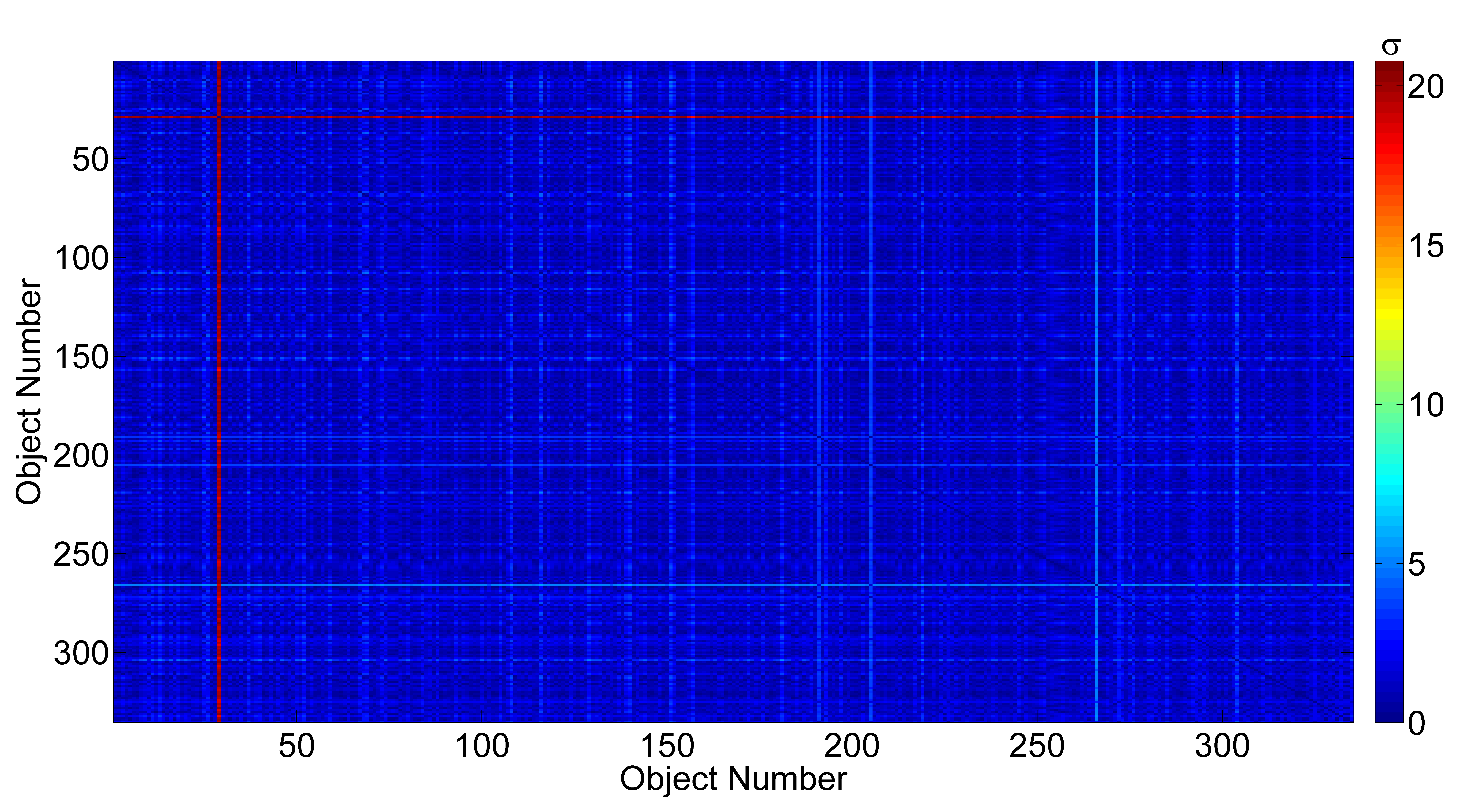}
   \caption{Color coded distance table of the objects in the set. A clear singular object is identified - object no. 29 - with a deviation on the order of 19$\upsigma$.}
   \label{2DMap}
\end{figure}

\begin{figure}[tb]
   \centering
   \includegraphics*[width=\linewidth]{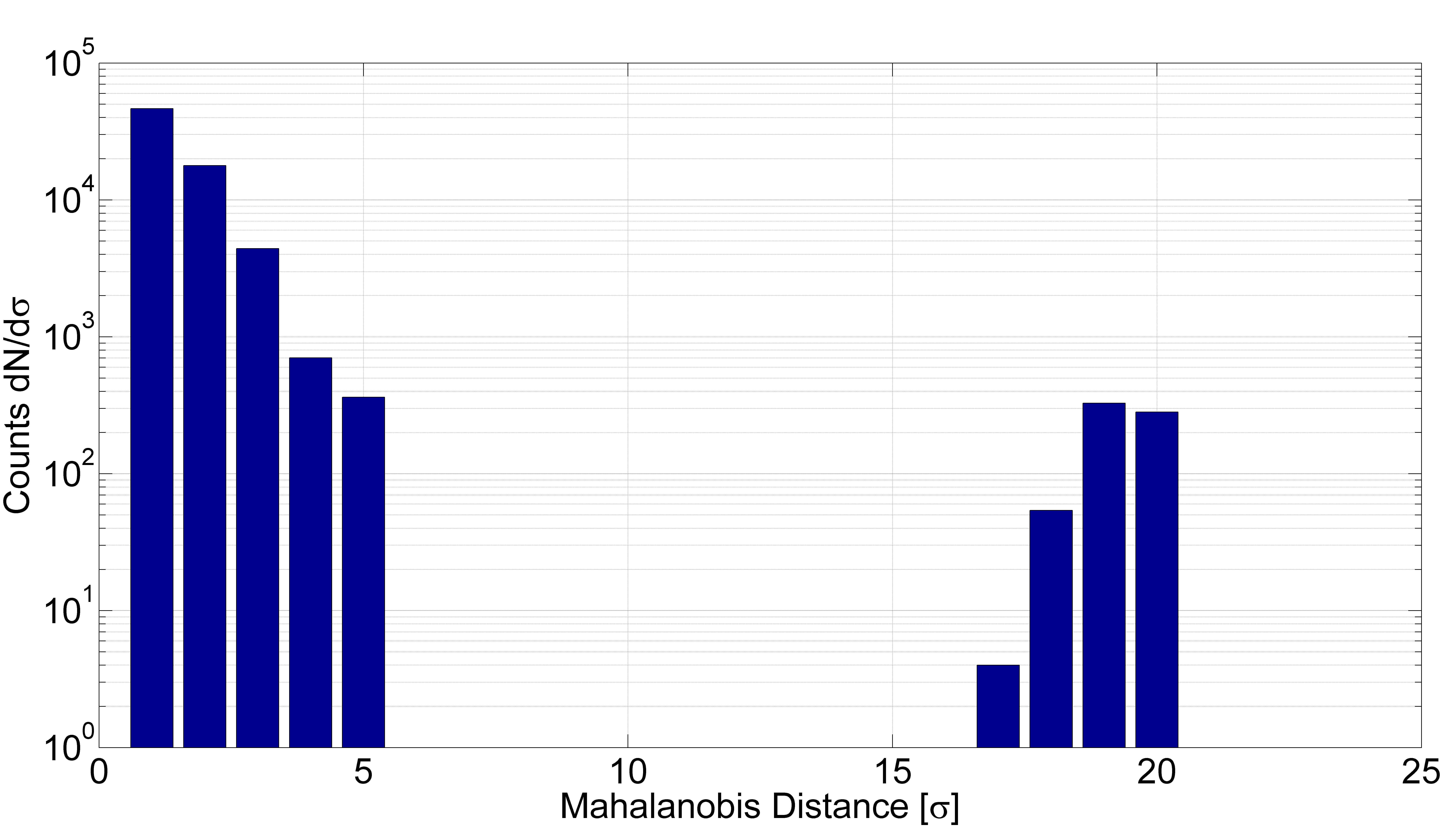}
   \caption{The histogram of the distance table. Object no. 29 - the defect - is a strong outlier of this set with an average deviation of 19$\upsigma$.}
   \label{HistMahal}
\end{figure}

The number of objects in the set surrounding the defect was 335. Figure \ref{2DMap} shows the distance table between these objects based on the Mahanalobis metric and Figure \ref{HistMahal} the histogram of the distance table.  

\begin{figure}[tb]
   \centering
   \includegraphics*[width=\linewidth]{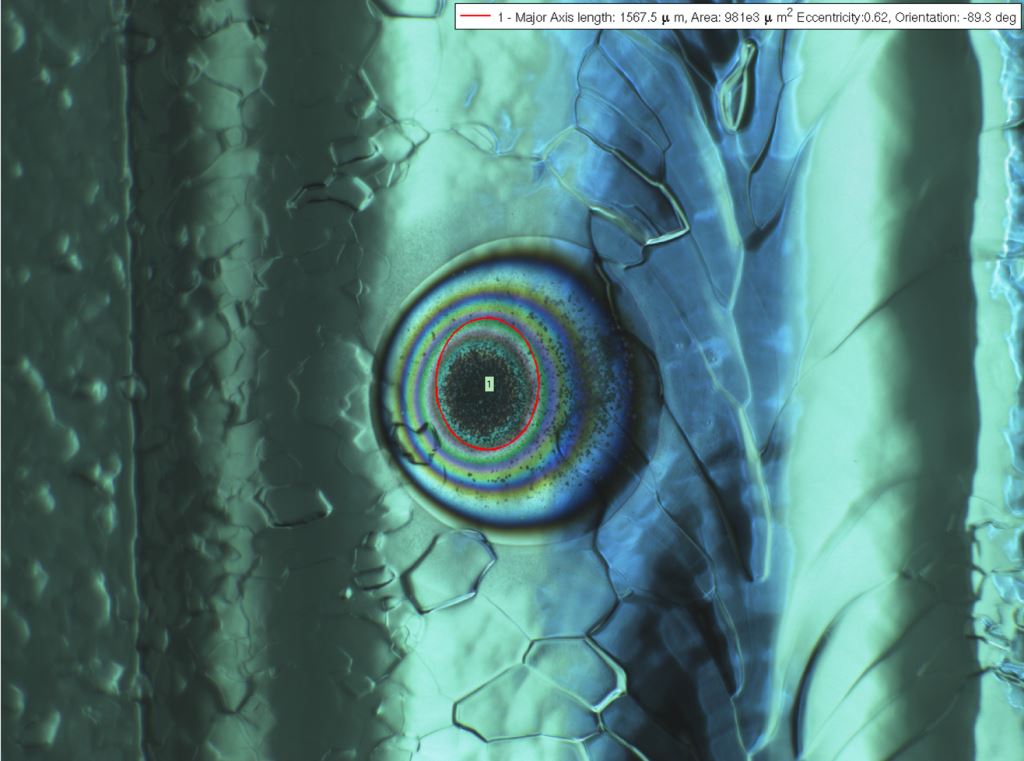}
   \caption{Oxidized niobium on a treated cavity surface after a failure in the high-pressure water rinsing leading to a reduction of the quench field. A short surface chemistry could remove the defect.}
   \label{Mahal_HPR}
\end{figure}

\begin{figure}[tb]
   \centering
   \includegraphics*[width=\linewidth]{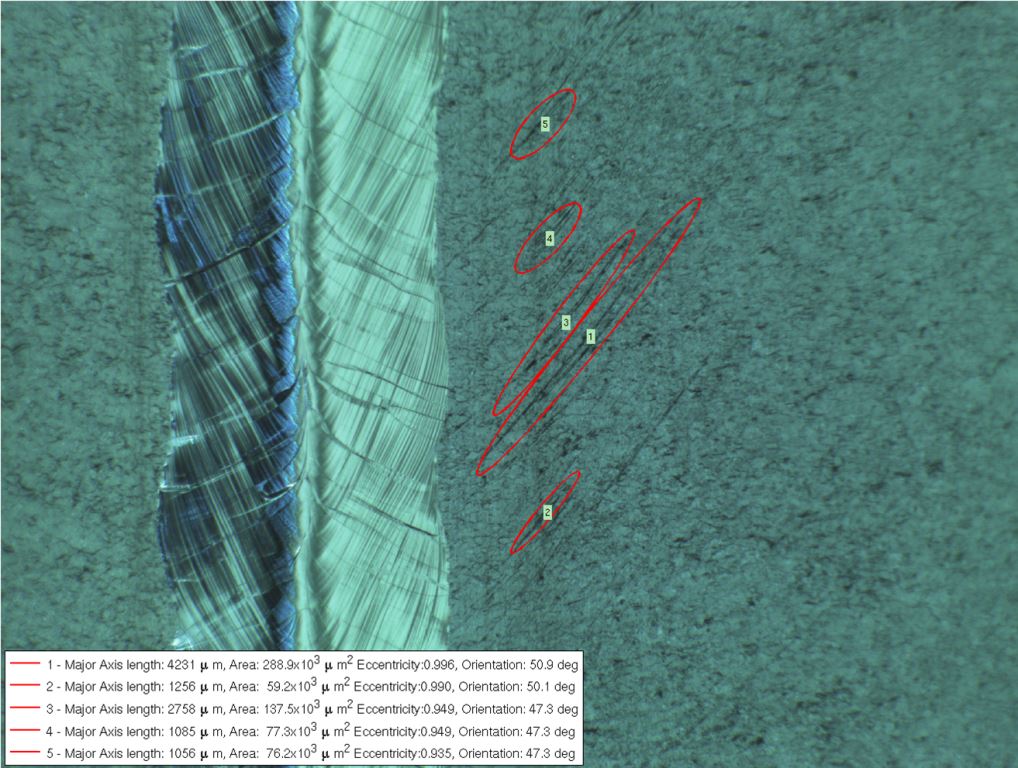}
   \caption{Scratches of unknown origin on the surface. The surface chemistry removed these defects and no further problems were observed.}
   \label{Mahal_Scratch}
\end{figure}

Using this approach, several other defects in other cavity inspections where automatically identified, see Figure \ref{Mahal_HPR} and \ref{Mahal_Scratch} as examples. 

A disadvantage of this classifier is the computing time. The algorithm ran on a server with two 3\,GHz Intel Core 2 DUO E8400 CPUs, 8\,GB memory and a GeForce 8400 with UBUNTU 14.04 as operating system. The image processing algorithm takes up to 40\,s CPU time per image to process it and to generate the phase space and additional 4-8\,min CPU time to scan the phase space for each object in the image while taking an image and move it in the computing infrastructure is on the order of 20\,s real time. Although each image can be processed parallel, a delay is unavoidable with given computational infrastructure, but this is acceptable since the overall delay is not time-critical.

\subsection{Eigenface Approach}
Another algorithm is based on the method of eigenfaces, which is the application of a principal component analysis to a collection of images to find a minimal set of eigenvectors - the so-called eigenfaces - which then can be used to describe and reconstruct the images. A variation has been implemented and used at OBACHT to recognize similarities between surface features \cite{Wenskat2012}. Consider $I\left(x,y\right)$ as the grey-scale intensity of a pixel at position $(x,y)$ of a square image $\Gamma_{i}$ consisting of $N\times N$ pixels. The same array can be interpreted as a vector with $N^{2}$ components, i.e. a point in a $N^2$-dimensional space. For a cavity surface area this space is not uniformly populated. In fact, the challenge of finding similarities between image areas is to determine the subspace in which the intensity only fluctuates statistically expected around the average gray scale value. This average value and the expected fluctuation should then only rely on the manufacturing variations of each vendor. Any deviation from this distribution should then be considered as a defect. The method applied is based on a training sample of $M$ cavity images $\Gamma_1,\Gamma_2,\ldots,\Gamma_M$.

The mean image $\bar{\Gamma}$ and its variance $C$ can be readily calculated
\begin{equation}\label{eq:mean}
    \bar{\Gamma}=\frac{1}{M}\sum^{M}_{i=1} \Gamma_i
\end{equation}
\begin{equation}\label{eq:covmatrix}
    C=\frac{1}{M}\sum^{M}_{i=1} (\Gamma_i- \bar{\Gamma})(\Gamma_{i}- \bar{\Gamma})^{T}
\end{equation}
which again represent arrays of size $N\times N$.
The eigenvectors of the matrix $C$ can be used to decompose the image space in terms of its most relevant properties, i.e. the eigenfaces. Typically only a small number of eigenvectors are needed to describe the relevant features of the image. The decomposition of a whole image into eigenfaces is not reasonable due to a simple problem: A whole image consists of welding seam region and heat affected zone region with distinct features and mixing these two would create a bi-model distribution complicating the defect identification. Hence, the training images were cut into multiple squares with an area of $101 \times 101$ pixel. The size of the squares is an important variable to reduce the false negative rate. A too large square could include a defect which does not attribute enough to the variance of this very square. A too small square would cut a defect into several squares where each might be classified as normal surface due to a too small variation in the square. Given the average diameter of a defect under consideration to be $35\,\upmu m$ (101 pixel) or bigger, this value was chosen, resulting in 34*26 squares to be analyzed per image. 

\begin{figure}[tb]
   \centering
   \includegraphics*[width=\linewidth]{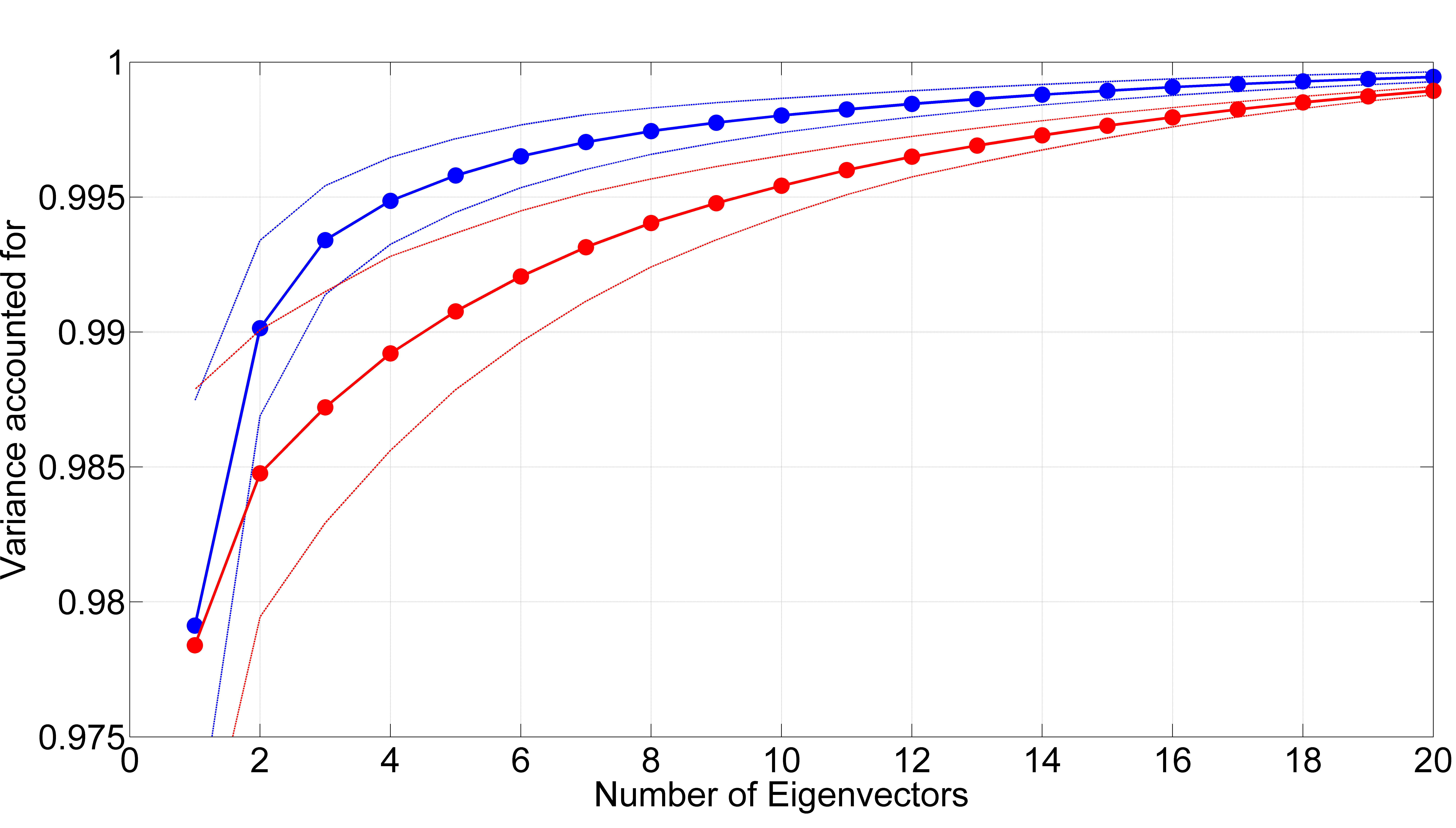}
   \caption{Variance accounted for in the image as a function of used eigenvectors for treated (red - 36 images) and untreated (blue - 23 images) surfaces.}
   \label{Eigenvectors}
\end{figure}

The selection of a minimal set of eigenvectors which accounts for the most variance in the image reduces the dimensionality of the search. Figure \ref{Eigenvectors} shows the variance accounted for as a function of eigenvectors used. On average, 10 eigenvectors are needed to account more than 99.5\% of the variance in the image for treated surfaces, while 5 eigenvectors are enough for untreated surfaces.   
Any irregularity of the surface, such as a geometrical defect or inclusion on the surface, deviates from the typical surface of the training sample and hence can be identified by distinctly different coordinates in eigenspace. Irregularities in the image are detected by introducing the Mahalanobis distance in the $N^2$-dimensional space. The threshold for detection was set to 5$\sigma$ and 10 eigenvectors are considered. 

\begin{figure}[tb]
   \centering
   \includegraphics*[width=\linewidth]{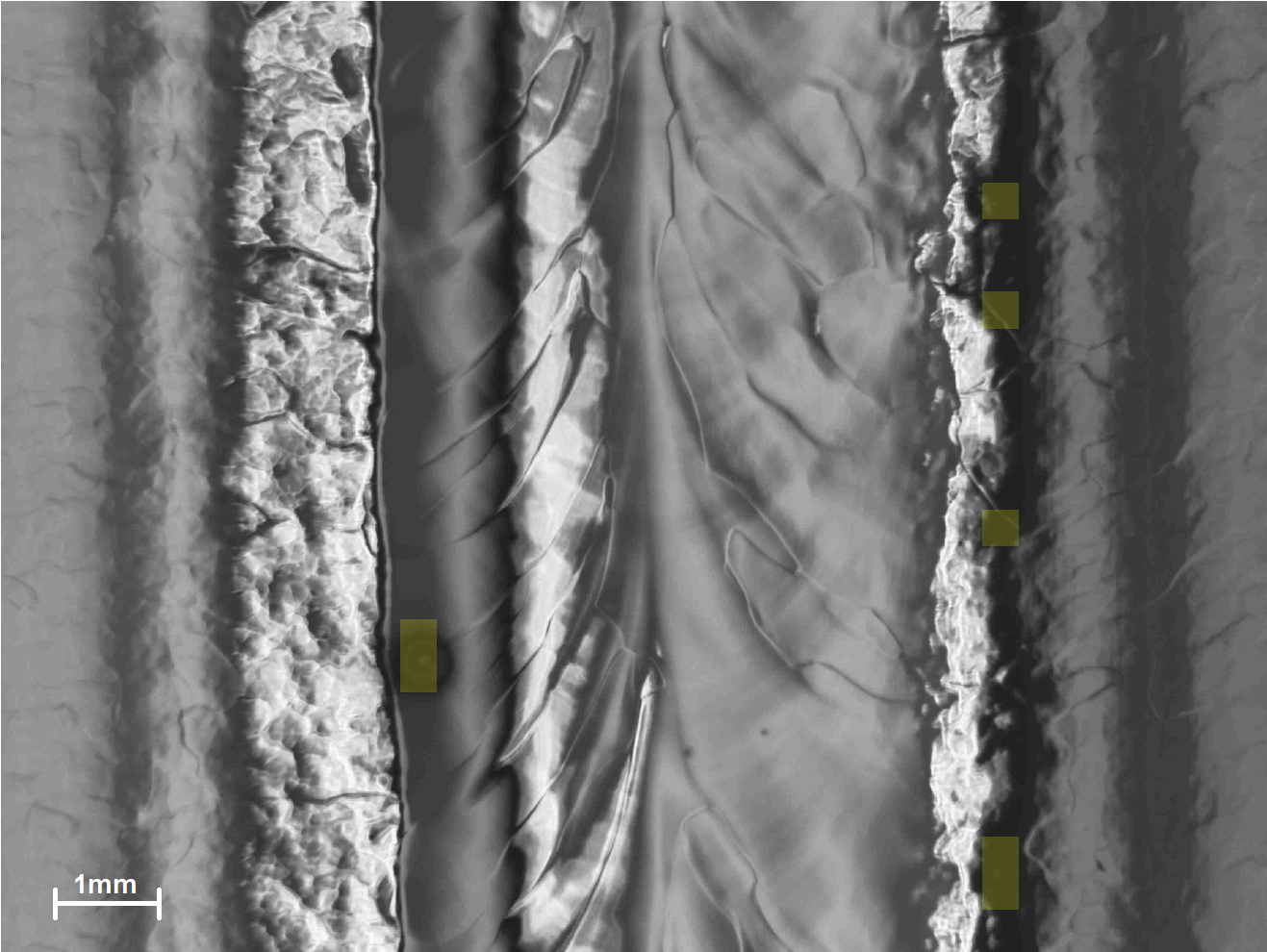}
   \caption{Gray scale image of the welding seam in the cavity. The output of the eigenfaces algorithm is indicated by the yellow squares. Two defects squares (two yellow squares on the left) have been identified, five other squares are false positives (shown as the 5 yellow squares on the right side of the welding seam).}
   \label{ac127}
\end{figure}

The algorithm has been tested on images of the optical inspection of a cavity. The rf defect could be localized and the optical inspection identified a spot which resembled a burst of debris. Figure \ref{ac127} shows the defect and detected squares.

\begin{figure}[tb]
   \centering
   \includegraphics*[width=\linewidth]{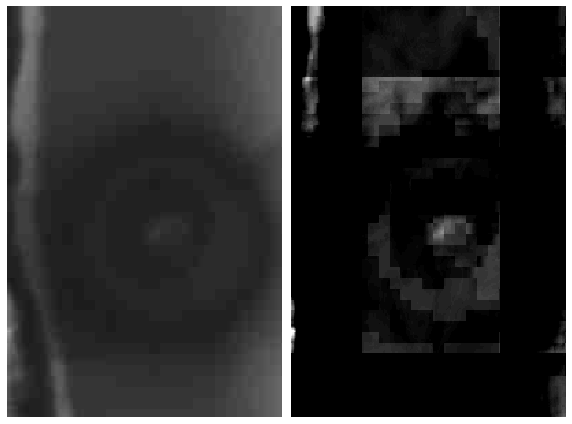}
   \caption{The original cross section of the defect area in the cavity (left) and after subtraction of the ten most relevant eigenvector weighted with the eigenvalues (right). Brighter regions are equivalent to a larger difference between actual image and reconstructed image.}
   \label{eigen}
\end{figure}

\begin{figure}[tb]
   \centering
   \includegraphics*[width=\linewidth]{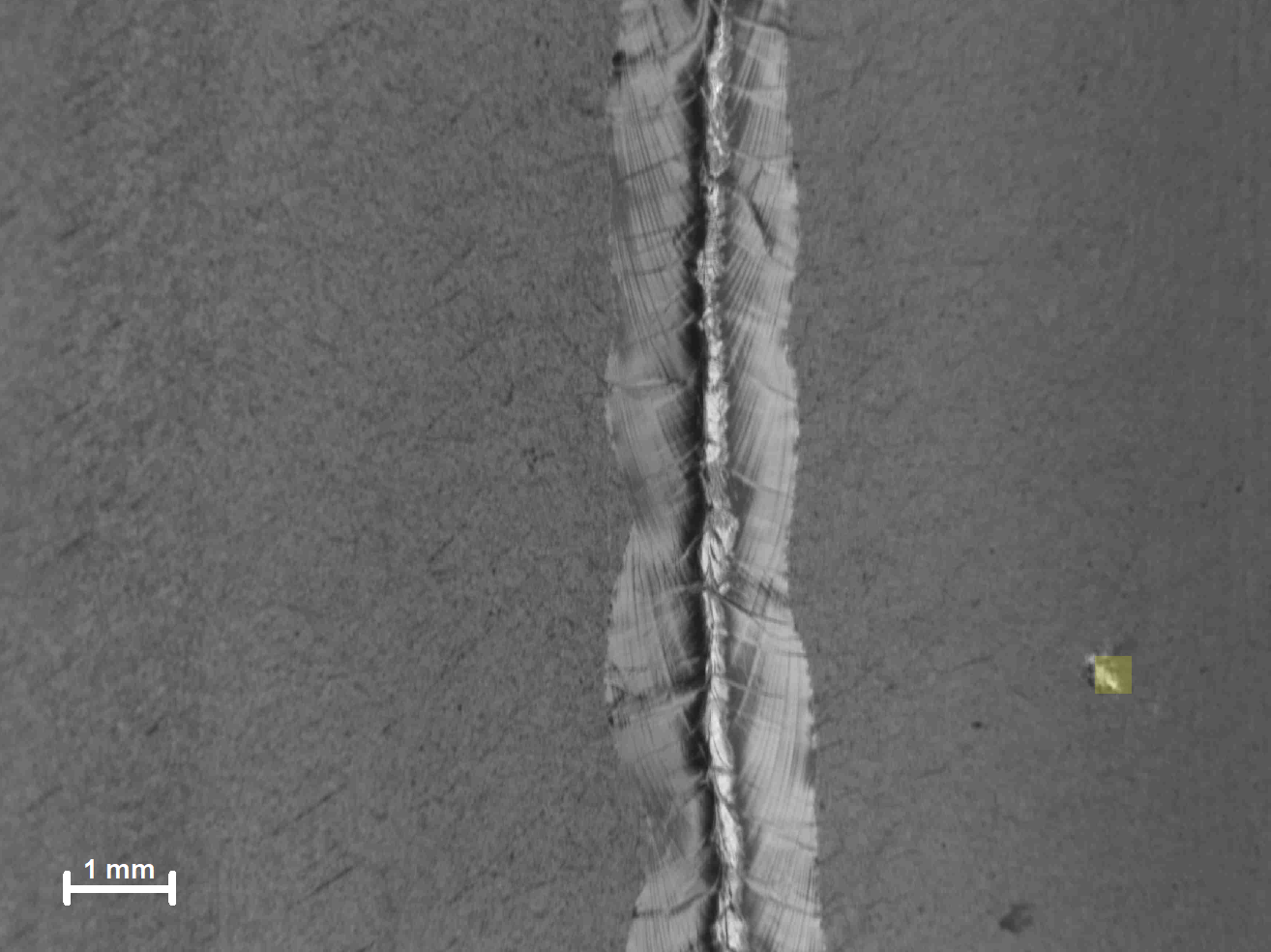}
   \caption{A welding splatter next to a welding seam of an untreated cavity surface - the square identified as defect marked in yellow.}
   \label{Eigen2}
\end{figure}

Figure \ref{eigen} on the left shows the defect and its direct neighborhood in the original image in gray scale while the right part shows the same region after subtracting the ten leading eigenvectors weighted with the appropriate eigenvalues. The signal to noise ratio (SNR) increases from 13.8~dB to 16.2~dB for the squares including the defect. At the same time, the SNR for the whole image increases by 0.7~dB indicating that most of the image was described by the background generated with the eigenfaces.
The false positive classification of the five squares could be traced to the chosen training set. All falsely classified squares contain the same light reflection pattern which was not well represented in the original training set. Figure \ref{Eigen2} shows another example of an image containing a defect identified with the Eigenface approach without any false positive squares.

A strong disadvantage of the eigenface method is that the algorithm identifies parts of the transient region of the seam in Figure \ref{ac127} as defective areas because it contains irregular reflective patterns. Increasing the amount of eigenvectors used or the training set does not decrease the false positive rate to a sufficient level since such reflective patterns are highly irregular. 

\section{Surface Classification}
Detecting defects and -as a next step - classifying automatically whether they can cause a limitation and at which accelerating field automatically is a difficult task. One issue is, that due to performance tests, only the 'worst' defect can be identified, while other limiting defects in the same cavity can remain undetected or not be classified from an RF point of view which reduces the available data set to maximum of one defect per cavity. With the optical inspection system OBACHT, a manual analysis was done during the European XFEL production \cite{Navitski2015b} and a classification of types of defects was done. But even cavities without a localized defect can quench at lower fields and first attempts in correlating global surface properties and not localized defects with performance parameters were done \cite{MW_SUST}. Here, an automated approach of a defect-free surface classification will be presented and is based on a classification tree \cite{Tree1}. A total of 33 cavities are used to train the tree. The limiting cell of each cavity was identified by RF tests and the surface properties of the respective cell and the quench field of the cavity are used. The quench fields are sorted into one of four classes, see Table \ref{tab:classestree}. Cells with a quench field below 30 MV/m would be rejected or need additional analysis and/or treatment in future accelerator projects (Class 1). Class 2 cells would be considered for light surface re-treatment. Class 3 cells are good cavities, while class 4 cells would be considered as excellent.

\begin{table}[ht]
	\centering
	\caption{Classification Table for 33 training cells. }
\begin{tabular}{lcc}
	\toprule
	\textbf{Class} & \textbf{$E_{acc}$ [MV/m]} & \#Cells \\ \cmidrule{2-3}
 	1 & $<$ 30 & 4 \\
	2 & $\leq$35 & 11 \\
	3 & $\leq$40 & 6 \\
	4 & $>$ 40 & 12 \\
 	\bottomrule
 	\end{tabular}

	\label{tab:classestree}
\end{table}

\begin{figure}[tb]
   \centering
   \includegraphics*[width=\linewidth]{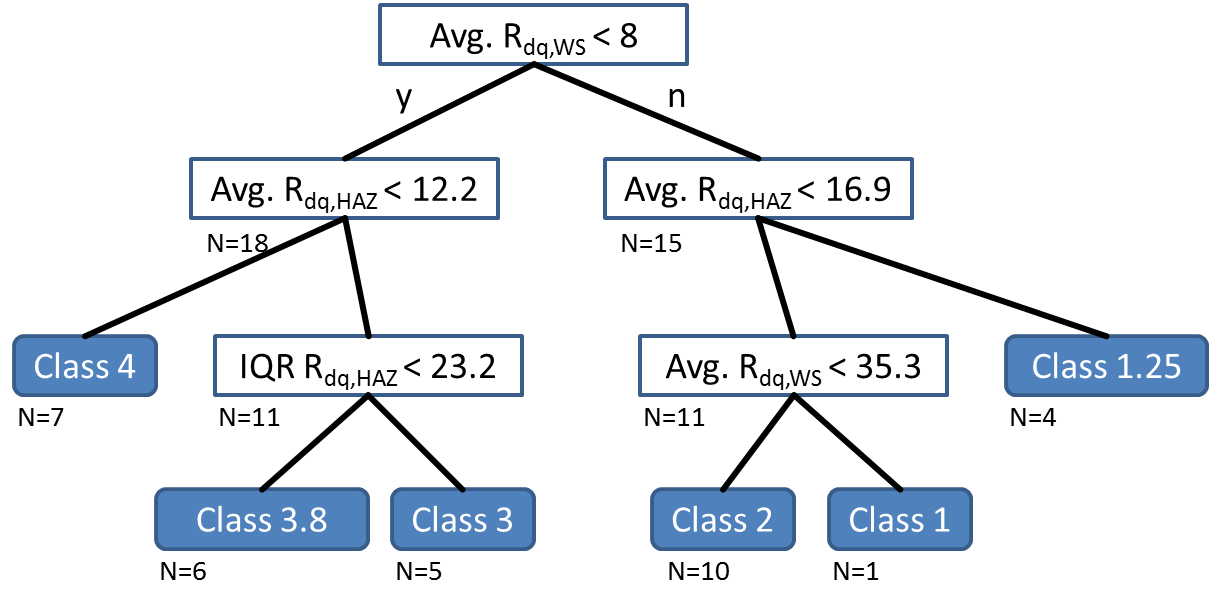}
   \caption{Binary tree after training with 33 cells.}
   \label{Treeplot}
\end{figure}

The different loss models discussed in the literature predict different correlations between the RF performance and the surface properties, but all of them assume the grain boundaries as a major potential for limitations or losses \cite{Visentin2003,Bauer2004,Ciovati2007,Ciovati2008a,hylton1988,Safa1999,Knobloch1999a}. Using the above mentioned variables obtained by the image processing algorithm, a set of 24 parameters for each cell were obtained. For example, the integrated grain boundary area $\sum{\mathrm{A}}$ in an image, but also for two regions - welding seam and heat affected zone - separately. Also, the average surface roughness $\overline{\mathrm{R_{dq}}}$ for a whole cell and for the regions and the spread of the surface roughness distribution - parametrized by the inter-quartile range - is used. These 24 parameters were used as predictors and the cell classes as response to be predicted. The full trained tree, shown in Figure \ref{Treeplot}, has five splits and two mixed classes in the terminal nodes.  
This decision tree is in agreement with the physical model, that smoother surfaces achieve higher fields \cite{Lilje} and that the welding seam is the most crucial part of the fabrication of a cavity. It is still surprising that the only variable used as predictor is the surface roughness in the different regions. A possible reason for this is the way the surface was represented in the variables. More sophisticated representations may improve the predictions and the surface roughness as the only important feature might be not correct anymore.

\section{Benchmark}
To quantify the classifiers, using common quantities as accuracy might skew the interpretation, since the two different object classes "defect" and "no defect" are populated highly asymmetrically - a situation called accuracy paradox. The object-oriented approach has on average one to three defects in an defect image and none in a defect free image, while it has on average 2000 grain boundaries. Using the eigenface approach, the image consists of 884 squares, where - depending on the defect size - one to twenty of those are part of a defect. To classify such asymmetric populations of classes, other metrics have been defined elsewhere and the g-mean and the F-measure will be used, where the range of these quantities are 1 (all correct classified) and 0 (nothing correct classified) \cite{Lewis1994,Kubat1997,Kubat1998,Fawcett2006}. The classifiers were tested on a set of 23 images of a treated cavity surface and 36 images of an untreated cavity surface, each containing at least one defect. The results of the training are given in table \ref{Table_Results}.
\begin{table}[ht]
	\centering
	\caption{Effectiveness of the two defect detection algorithms for treated and untreated surface images}
\begin{tabular}{lcccc}
	\toprule
	 \multicolumn{1}{c}{}           &    \multicolumn{2}{c}{\textbf{Eigenface} } &    \multicolumn{2}{c}{\textbf{Object-Oriented}}    \\
		& Treated & Untreated & Treated & Untreated \\
	\cmidrule{2-5}
 	$\mathrm{g-mean_1}$ & 0.53 & 0.63 & 0.37 & 0.86 \\
	$\mathrm{g-mean_2}$ & 0.88 & 0.89 & 0.70 & 0.99 \\
	F-measure & 0.50 & 0.61 & 0.36 & 0.85 \\
	\bottomrule
 	\end{tabular}
	\label{Table_Results}
\end{table}
In addition, the eigenface algorithm didn't detect any defect in 13 of 23 treated and 10 of 36 untreated surface images and the object-oriented algorithm didn't detect any defect in 7 of 23 treated and 6 of 36 untreated surface images and hence the values are zero for those images.  

\begin{figure}[tb]
   \centering
   \includegraphics*[width=\linewidth]{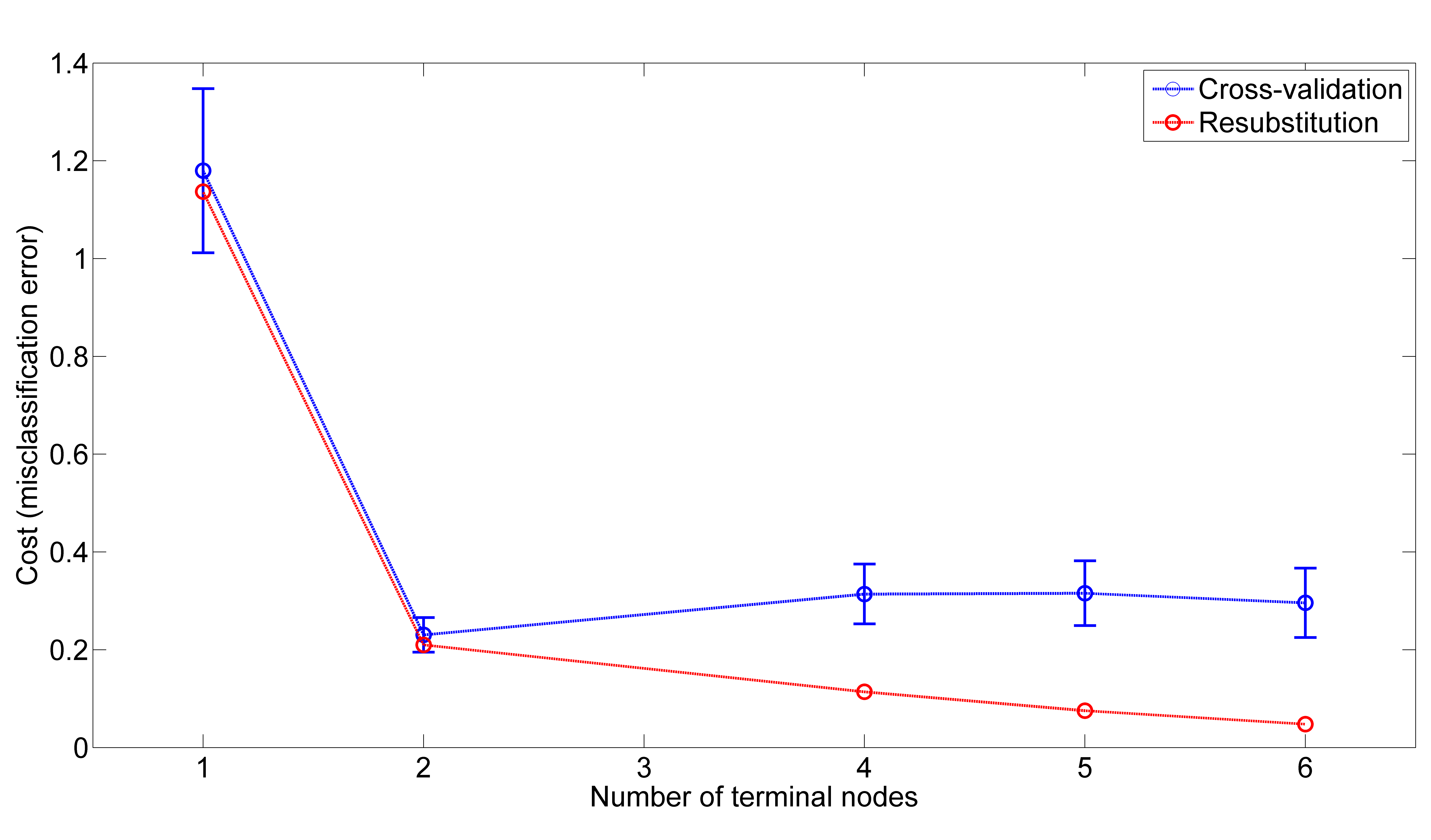}
   \caption{The missclassification error at a node against the number of terminal nodes. Two methods are shown, to estimate the missclassification error.}
   \label{Treetest}
\end{figure}

Figure \ref{Treetest} shows the cost of the tree using a resubstitution method and a cross-validation method. The cost of a node is the sum of the misclassification costs of the observations in that node. The resubstitution cost is based on the same sample that was used to create the original tree, so it under estimates the likely cost of applying the tree to new data. The cross-validation method uses a 10-fold cross-validation to compute the cost vector. The function partitions the sample into 10 subsamples, chosen randomly but with roughly equal size and class proportions.
A misclassification of (35 $\pm$ 7) \% is expected. The decision tree was then applied to a another set of 17 cells produced within the European XFEL fabrication and which were found to be the limiting cells in cavities and are defect free \cite{MW_SUST}. The results is shown in a confusion table in Table \ref{tab:classestree_apply}.  
\begin{table}[ht]
	\centering
	\caption{Confusion Table for 17 cells.}

\begin{tabular}{lcccc}
	\toprule
	\textbf{Class} & 1 & 2 & 3 & 4 \\ \cmidrule{2-5}
 	1 & 4 & 0 & 0 & 0 \\
	2 & 0 & 4 & 1 & 1 \\
	3 & 0 & 3 & 2 & 0 \\
	4 & 0 & 0 & 1 & 1 \\
 	\bottomrule
 	\end{tabular}
	\label{tab:classestree_apply}
\end{table}
35\% of the cells were misclassified, which is in excellent agreement with the expected value. More important, no class 1 cavities were misclassified or cavities were falsely classified as class 1. 

\section{Conclusion}
For the implementation of the optical inspection robot OBACHT as a quality assurance and control tool in a large scale production, as well as an R\&D tool, an automated image processing and analysis algorithm was needed. In the scope of the ILC HiGrade Research Project and the European XFEL cavity fabrication, a new framework has been developed which enables this automated analysis of a large amount of images of the inner surface of cavities \cite{Wenskat2017}. First applications of the newly developed framework were investigation of optical surface properties of the two cavity vendors for the European XFEL and significant differences in the quantitative characterization have been identified and a standard for a cavity surface has been established \cite{MW_SUST}. The next steps, and first results as shown in this work, are to detect defects on an irregular surface automatically and classify their potential limitations as well as to classify the "goodness" of a defect free cavity surface. Two algorithms have been applied with moderate results. The object oriented approach is more promising and will be combined with other classifiers to further study its accuracy. The Eigenface approach shows high false positive and false negative rates on treated cavity surfaces. For untreated cavities it should be further tested - which is not possible anymore at the current state of the fabrication process. The classification of defect free surfaces achieved an already sufficient level of accuracy with a simple decision tree algorithm and could be used as a first level quality control tool during early fabrication steps. 
In conclusion, a quantitative analysis and characterization of a cavity surface by means of optical methods has been achieved, which can be adapted and used for the quality assurance of a future large scale cavity production.

\section*{Acknowledgment}
The author would like to thank S. Aderhold (now SLAC), A. Navitski (now RI), J. Schaffran (DESY), L. Steder (DESY) and Y. Tamashevich (now HZB) for their work. Otherwise, mine would have been impossible. This work is funded from the EU 7th Framework Program (FP7/2007-2013) under grant agreement number 283745 (CRISP) and "Construction of New Infrastructures - Preparatory Phase", ILC-HiGrade, contract number 206711, BMBF project  05H12GU9, and from the Alexander von Humboldt Foundation

\end{document}